\documentclass[twocolumn,eqsecnum,aps]{revtex4}

\usepackage{graphicx}
\usepackage{bm}
\usepackage{amssymb}
\usepackage[centerlast]{subfigure}

\newcommand{\zset}{{\mathbb{Z}}}
\newcommand{\rset}{{\mathbb{R}}}

\newcommand{\ra}{\right \rangle}
\newcommand{\la}{\left \langle}

\begin{document}

\title{Forced Burgers Equation in an Unbounded Domain}
\author{J\'er\'emie Bec} \email{bec@obs-nice.fr}
\affiliation{Observatoire de la C\^ote d'Azur, Lab.\ G.-D.\ Cassini, BP
  4229, Nice Cedex 4, France.\\
  Institute for Advanced Study, Einstein Drive, Princeton, NJ 08540, USA.}
\author{Konstantin Khanin} \email{kk262@newton.cam.ac.uk}
\affiliation{Isaac Newton Institute for Mathematical Sciences, 20
  Clarkson Road, Cambridge CB30EH, UK.\\ Department of Mathematics,
  Heriot-Watt University, Edinburgh EH144AS, UK.\\ Landau Institute
  for Theoretical Physics, Kosygina Str.,2, Moscow 117332, Russia.}
\draft{\textit{J.\ Stat.\ Phys.}, in press (2003)}

\begin{abstract}
  The inviscid Burgers equation with random and spatially smooth
  forcing is considered in the limit when the size of the system tends
  to infinity.  For the one-dimensional problem, it is shown both
  theoretically and numerically that many of the features of the
  space-periodic case carry over to infinite domains as intermediate
  time asymptotics. In particular, for large time $T$ we introduce the
  concept of $T$-global shocks replacing the notion of main shock
  which was considered earlier in the periodic case (1997, E et al.,
  \textit{Phys.\ Rev.\ Lett.}\/ \textbf{78}, 1904). In the case of
  spatially extended systems these objects are no anymore global.
  They can be defined only for a given time scale and their spatial
  density behaves as $\rho(T)\sim
  T^{-2/3}$ for large $T$. The probability density function $p(A)$ of
  the age $A$ of shocks behaves asymptotically as $A^{-5/3}$.  We also
  suggest a simple statistical model for the dynamics and interaction
  of shocks and discuss an analogy with the problem of distribution of
  instability islands for a simple first-order stochastic differential
  equation.\\[3pt]
\textbf{Key words:} Burgers turbulence, stochastic forcing, shock
discontinuities
\end{abstract}

\maketitle

\section{Introduction}
\label{sec:intro}
\noindent The $d$-dimensional forced Burgers equation
\begin{equation}
  \partial_t {\bm u} +  \left({\bm u}\cdot\nabla\right){\bm u} = \nu
  \nabla^2 {\bm u} - \nabla F({\bm x},t), \qquad {\bm u} =
  -\nabla \psi
  \label{burgers_dim_d}
\end{equation}
which describes the dynamics of a stirred, pressure less and vorticity-free
fluid, has found interesting applications in a wide range of non-equilibrium
statistical physics problems. It arises, for instance, in cosmology where it
is known as the adhesion model \cite{gss89}, in vehicular traffic \cite{css00}
or in the study of directed polymers in random media \cite{bmp95}. The
associated Hamilton--Jacobi equation, satisfied by the velocity potential
$\psi$,
\begin{equation}
  \partial_t \psi -\frac{1}{2} \left | \nabla \psi \right |^2 =
  \nu \nabla^2 \psi + F({\bm x},t),
  \label{kpz}
\end{equation}
has been frequently studied as a nonlinear model for the motion of an
interface under deposition\,: when the forcing potential $F$ is
random, delta-correlated in both space and time, Eq.~(\ref{kpz}) is
the well-known Kardar--Parisi--Zhang (KPZ) equation~\cite{kpz86}. The
case with large-scale forcing was considered in Refs.~\cite{s91,cy95}
as a natural way to pump energy in order to maintain a statistical
steady state. Burgers equation is then a simple model for studying the
influence of well-understood structures (shocks, preshocks, etc) on
the statistical properties of the flow. As it is well known,
Eq.~(\ref{burgers_dim_d}) in the limit of vanishing viscosity
($\nu\to0$) displays after a finite time dissipative singularities,
namely shocks, corresponding to discontinuities in the velocity field.
In the presence of large-scale forcing, it was recently stressed for
the one-dimensional case \cite{ekms97,ekms00} and also for higher
dimensions \cite{ik01,bik02}, that the global topological structure
of such singularities is strongly related to the boundary conditions
associated to the equation. More precisely, when, for instance, space
periodicity is assumed, a generic topological shock structure can be
outlined. It plays an essential role in understanding the qualitative
features of the statistically stationary regime.

So far, the singular structure of the forced Burgers equation was
mostly investigated in the case of finite-size systems with periodic
boundary conditions. It is however frequently of physical interest to
investigate instances where the size of the domain is much larger than
the scale, so as to examine, for example, the role of Galilean
invariance \cite{p95}. The main goal of the present paper is to
describe the singular structure of Burgers
equation~(\ref{burgers_dim_d}) in unbounded domains. To achieve this
goal we consider the case of spatially periodic forcing with a forcing
scale $L_f$ much smaller than the system size $L$.  More
precisely, we assume that the force is random and white in time,
homogeneous, isotropic, smooth and $L$-periodic in space and
investigate the limit $L/L_f \gg 1$, while keeping a constant
injection rate of energy. Hence, for a fixed size of a system $L$ we
consider a stationary regime corresponding to the limit $t \to \infty$
and then study the limit $L \to \infty$ with the ${\cal L}^2$ norm of
the forcing growing like $L$. As we will see, most of the concepts
associated to infinite time asymptotics introduced for the periodic
case can be generalized in this limit as intermediate time asymptotics.

It is important to mention that the problem we are discussing has also
another interpretation which has no direct relation to Burgers
equation.  In fact, we are studying the structure of singularities for
variational problems associated to generic time-dependent Lagrangians
in unbounded domains. From this point of view the random forcing is
just a natural way to characterize generic time-dependence.

The paper is organized as follows.  In Sec.\ \ref{sec:periodic} we give a
brief exposition of the theory in the case of periodic forcing.  In
particular, we introduce the variational approach to Burgers turbulence and
present results on the existence and the uniqueness of the main shock and the
global minimizer. In Sec.\ \ref{sec:tscale} we introduce the notion of
$T$-global shocks and describe their behavior in the time asymptotics
$L_f/u_{\rm rms}\ll T\ll L/u_{\rm rms}$, where $u_{\rm rms}$ is the
root-mean-square velocity. We also discuss theoretical and numerical results
for the density of $T$-global shocks and the probability density function
(PDF) of the age of shocks.  In Sec.\ \ref{sec:models}, we discuss an analogy
with stability theory for one-dimensional stochastic ordinary differential
equation (ODE) and present a simple statistical model for the interaction of
shocks.  Section \ref{sec:conclusion} contains concluding remarks and, in
particular, an interpretation of the results in terms of an inverse cascade.
Most of the results of this paper are focusing on the one-dimensional case. We
finish with a brief discussion of possible extensions to the multi-dimensional
case.

\section{Periodic forcing}
\label{sec:periodic}
\noindent In the case of smooth spatially periodic forcing potential
$F({\bm x},t)$, it was shown in the one-dimensional case \cite{ekms00}
and for higher dimensions \cite{ik01} that a statistical steady state
is reached at large times by the solution to the Burgers equation.
Taking the initial time at $-\infty$, the velocity field is periodic
in space and uniquely determined by the realization of the forcing. At
an arbitrary time $t$, this unique solution can be expressed in terms
of the following variational principle \cite{l57,o57,l82} involving
the Lagrangian action $\cal A$\,:
\begin{eqnarray}
  {\bm u}({\bm x},t) &=& \nabla_{\bm x} \min_{{\bm \gamma}:{\bm
    \gamma}(t)=x} \left [ {\cal A} \left  ( \bm\gamma,t\right) \right ],
  \label{minimum} \\ {\cal A} \left ( \bm\gamma,t\right) &\equiv &
  \int_{-\infty}^{t} \left[ {1\over2} \left| \dot{\bm \gamma}
      (\tau)\right|^2 - F\left({\bm \gamma}(\tau),\tau \right)\right] d\tau.
  \label{defaction}
\end{eqnarray}
The minimum in (\ref{minimum}) is taken over all absolutely continuous curves
${\bm \gamma}(\tau)$ of $\rset^d$ with $\tau\in (-\infty,t]$ such that ${\bm
  \gamma}(t)={\bm x}$.  A curve minimizing the action in~(\ref{minimum}) is
called a {\em one-sided minimizer}.  It corresponds to a fluid particle
trajectory which is not absorbed by shocks till time $t$.  For all times
$\tau<t$, such a minimizer is a trajectory of the dynamical system
corresponding to the Lagrangian flow and thus satisfies the Newton equation
\begin{equation}
  \ddot{\bm \gamma}(\tau) = - \nabla F ({\bm \gamma} (\tau),\tau).
  \label{newton}
\end{equation}
Since the force is potential and spatially periodic, the mean velocity
\begin{equation}
  \bm b \equiv \frac{1}{L^d}\int_{[0,L]^d} \bm u({\bm x},t) d{\bm x}
  \label{meanvel}
\end{equation}
is the first integral of (\ref{burgers_dim_d}).  When the initial time
is taken at $-\infty$, so that the statistically stationary regime is
reached, the mean velocity $\bm b$ is the only information remaining
from the initial condition. For simplicity, we consider in the sequel
the case of a vanishing mean velocity ($\bm b=0$), but all the results
remain true for arbitrary~$\bm b$.

\begin{figure}[htbp]
  \centerline{\includegraphics[width=0.35\textwidth]{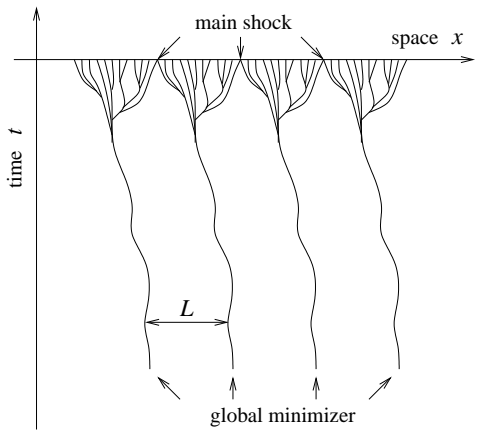}}
  \caption{Sketch of the unwrapped picture in space-time for $d=1$; each
    minimizer converges backward in time to one of the periodic image
    of the global minimizer. The main shock is defined as the location
    from which emanate two minimizers approaching different images of
    the global minimizer.}
  \label{unwrap1d}
\end{figure}

It is easy to show that for Lebesgue almost all $\bm x$ there exists a
unique one-sided minimizer.  The locations where there are more than
one minimizer correspond to shock positions. Those are exactly the
positions where the velocity potential $\psi({\bm x},t)$, which is a
Lipschitz function, is not differentiable, so that the piecewise
continuous velocity field ${\bm u}({\bm x},t)=-\nabla \psi({\bm x},t)$
has jump discontinuities.  As $t \to-\infty$, all the one-sided
minimizers which originated at time $t$ converge backward-in-time to
the trajectory of the unique fluid particle which is never absorbed by
a shock. This trajectory, denoted $\bm \gamma_{\rm gm}(\cdot)$, is
called the {\em global minimizer} because it is an action-minimizing
trajectory at any time. The statement about uniqueness of the global
minimizer holds under some natural conditions on the forcing potential
(see, e.g., Refs.\ \cite{ekms00,ik01}). Shocks are generically born at
some finite time and then grow and merge. There exist a particular
shock structure, called the {\em main shock} ($d=1$) or the {\em
topological shock} ($d>1$) which has always existed in the past.  This
shock can be constructed by unwrapping the picture to the whole space
$\rset^d$ (see Fig.\ \ref{unwrap1d} for the case $d=1$). We then
obtain a lattice of periodic boxes, each of them containing a periodic
image of the global minimizer and the {\em topological shock}
corresponds to the set of ${\bm x}$-positions giving rise to several
minimizers that approach different global minimizers belonging to
different images of the periodic box.

We focus now on the one-dimensional case with a space-periodic forcing
potential of period $L$. In this case, the global minimizer is a
hyperbolic trajectory of (\ref{newton}) and is thus associated to a
smooth unstable manifold which is a smooth curve in the
position-velocity phase-space $(x,u)$.  The main shock is defined as
the unique position giving rise to the left-most and the right-most
one-sided minimizers approaching the global one backward-in-time. By
definition, the trajectory of the global minimizer $\gamma_{\rm
gm}(t)$ and the main shock trajectory $X_{\rm ms}(t)$ never intersect,
so that they have to satisfy $X_{\rm ms}(t) < \gamma_{\rm gm}(t) <
X_{\rm ms}(t)+L$. Hence, the large-scale displacement of the two-sided
minimizer is the same as for the main shock.  The Lagrangian
flow~(\ref{newton}) is a second-order stochastic ODE.  Assuming the
vanishing of the mean velocity $b$, one may be tempted to think that
the displacement of a typical trajectory $\gamma(t)$ scales like the
integral of the Brownian motion\,:
\begin{equation}
  \la \left ( \gamma(t)-\gamma(0) \right ) ^2 \ra \propto | t |^{3}
  \quad\mbox{when } | t | \to \infty.
  \label{typic-newton}
\end{equation}
However this scaling does not apply to the minimizers because the
particularity of such trajectories cannot be ignored. Indeed, when
investigating the typical displacement of the global minimizer, we
find that
\begin{equation}
  \la \left ( \gamma_{\rm gm}(t) - \gamma_{\rm gm}(0) \right ) ^2 \ra
  \propto |t| \quad\mbox{when } | t | \to \infty.
  \label{dev-glob-mini}
\end{equation}
This result was obtained by numerical study of the behavior at large
times of the position of the main shock. This gives the large-time
typical displacement of the position of the main shock which, as we
have seen above, coincides up to a constant with the displacement of
the global minimizer.
\begin{figure}[htbp]
  \centerline{\includegraphics[width=0.35\textwidth]{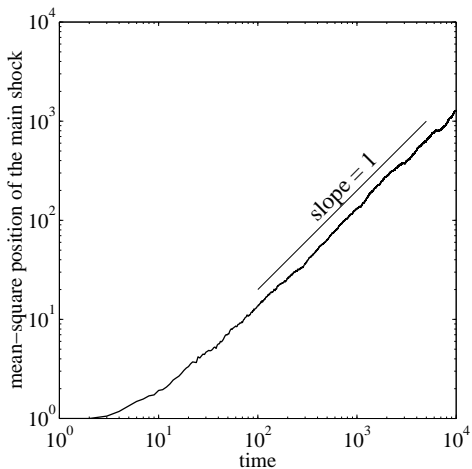}}
  \caption{Average over 500 realizations of the position $X(t)$ of the
    main shock as a function of time for $b=0$ and a space-periodic forcing
    with period $L=L_f$.}
  \label{f:devglobmini}
\end{figure}
For the numerical investigation, we chose a forcing which is the sum
of independent random impulses acting at discrete ``kicking'' times
\cite{bfk00}. Between successive kicks, the solution to the Burgers
equation decays.  This method is particularly efficient for obtaining
the solution in the limit of vanishing viscosity, as one can use
between kicks the fast Legendre transform method \cite{nv94} which
is directly connected with the variational principle (\ref{minimum})
and the Lagrangian picture of the flow. The position of the main shock
is obtained by a Lagrangian method which is based on the analysis of
the right-most and the left-most positions from which the two-sided
minimizer is approached.  Figure~\ref{f:devglobmini} shows the
diffusive-like behavior of the main shock trajectory for a Gaussian
forcing confined to the first three Fourier modes in the case when the
mean velocity $b$ vanishes.

The diffusive behavior of the global minimizer has also a simple
theoretical explanation.  Indeed, the mean square displacement of the
global minimizer can be expressed through its velocity as
\begin{equation}
  \la \left ( \gamma_{\rm gm}(t) - \gamma_{\rm gm}(0) \right ) ^2 \ra
  = \la \left ( \int_0^t \dot \gamma_{\rm gm}(s)\,ds \right ) ^2 \ra ,
  \label{dev-glob-mini1}
\end{equation}
where the velocity of the global minimizer $\dot\gamma_{\rm gm}(s)$ is
a stationary random process.  The hyperbolicity of the global minimizer
implies that the time correlations for this process decay
exponentially.  Hence, there are just two possibilities: the
mean-square displacement (\ref{dev-glob-mini1}) either grows linearly
with time, which gives (\ref {dev-glob-mini}), or it stays bounded in
a limit $t \to \infty$.  The latter behavior can be easily ruled out
since the trajectory of the global minimizer does fluctuate.

\section{Statistics of $T$-scales in large-size systems and PDF for
  the age of shocks.}
\label{sec:tscale}
\noindent As we have discussed above, the existence of the main shock
in the spatially periodic situation follows from a simple topological
argument.  The main shock is unique when the system is considered on
the torus corresponding to its period (circle for $d=1$). However, if
the system is considered on the whole real line (universal cover),
then there exists, at each moment of time, an $L$-periodic
one-dimensional lattice of main shocks.  Each of these shocks is
associated to two minimizers (left-most and right-most) which also
form a periodic structure.  In the case of the main shocks the
distance between these two minimizers tends to $L$ as $t\to-\infty$.
For all other shocks the corresponding distance tends to 0. In this
sense, the main shocks are the only global shocks present in the
system. There is also another way to characterize the main shock.
Namely, the main shock is the only shock which existed forever in the
past, that is, the main shock is infinitely old, contrary to all other
(local) shocks, all of them being created at a finite time and hence
having a finite age. In other words, local shocks can be traced
backward in time only for a certain finite interval of time. The
length of this time interval is what we call the age $A$ of the shock.

When we drop the periodicity condition and consider large-size
systems, the main shock disappears. But one can still introduce the
notion of \emph{$T$-global shocks}, that is shocks which behave like
the main shock when observed during a time scale of the order of
$T$. As in the case of periodic boundary conditions, there are now
also two equivalent ways to characterize $T$-global shock.
Geometrically, one can again consider left-most and right-most
minimizers associated to a given shock. After tracing them backward in
time for time intervals longer than a certain ``correlation time'',
these two minimizers are getting close and start converging to each
other exponentially fast.  $T$-global shocks are defined as shocks for
which this ``correlation time'' is larger than $T$.  This means that
for backward times $t \ll T$, one has $d(t)\gtrsim L_f$, where $d(t)$
is the distance between minimizers.  Equivalently, one can say that
$T$-global shocks have a finite age larger than $T$. Below, we study
the statistics of the $T$-global shocks.

We start with a numerical investigation of the forced Burgers equation
in a domain of a size $L$, much larger than the forcing length-scale
$L_f$.  The force is taken to be Gaussian, statistically homogeneous,
white-noise in time and characterized by its covariance
\begin{equation}
  \langle f(x+\ell,t+\tau) f(x,t) \rangle = B(\ell) \, \delta(\tau),
  \label{covarforce}
\end{equation}
where $B$ is a smooth, $L$-periodic function.  $L_f$ is the
correlation length, chosen such that $B(\ell)$ is concentrated at
scales $\ell<L_f\ll L$. The injection rate of energy per unit length
is kept constant (i.e.\ we choose the space correlation such that
$B(0)\propto L$). Such a hypothesis is essential because we found that
it prevents any $L$ dependency of the root-mean-square velocity
$u_{\rm rms}\equiv\la u^2(x,t) \ra^{1/2}$. We will now study the
behavior of the statistically stationary solution for the intermediate
time asymptotics $L_f/u_{\rm rms} \ll t \ll L/u_{\rm rms}$.  For very
large but finite $L$, the problem is well-defined and we can suppose
that the initial condition is taken at $-\infty$, so that a
statistical steady state is achieved for the solution of the Burgers
equation.

\begin{figure}[htbp]
  \subfigure[\label{f:velocity}]
  {\includegraphics[width=0.5\textwidth]{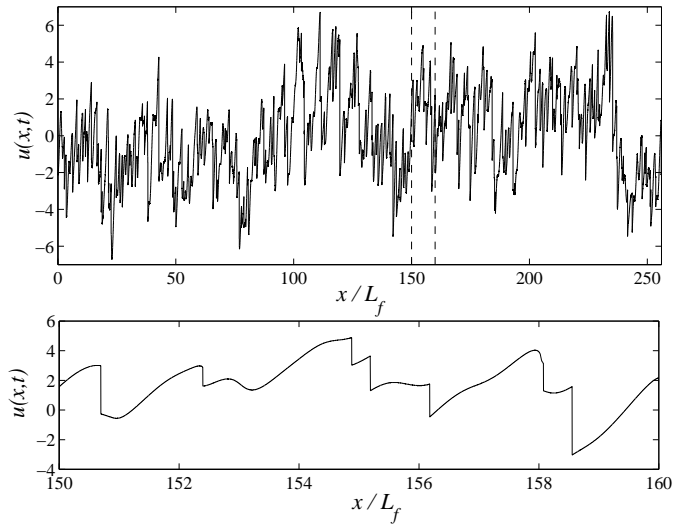}}
  \hfill
  \subfigure[\label{f:mini}]
  {\includegraphics[width=0.4\textwidth]{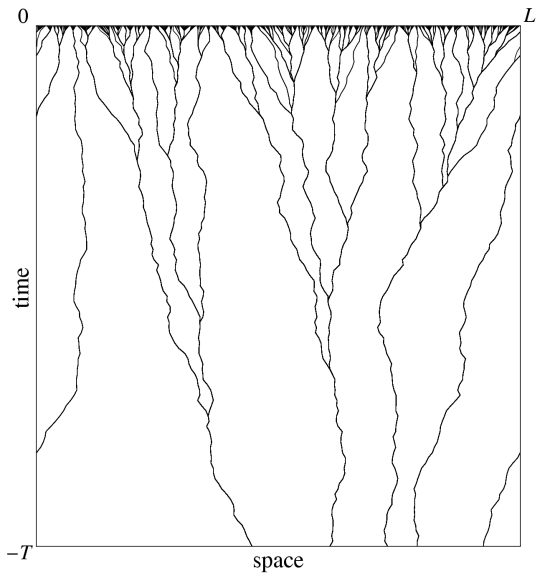}}
  \caption{(a) Upper: snapshot of the velocity field for
    $L=256L_f$. Lower: zoom of the field in an interval of length
    $10L_f$, represented by dashed lines in the upper figure. (b)
    Minimizing trajectories in space time for the solution to the
    randomly forced Burgers equation periodic in space with period
    $L=256 L_f$ and over a time interval of length $T=100$. The
    various trajectories are chosen sufficiently separated at time
    $t=0$ in order to lighten the picture. The minimizers converge to
    each other in a rather nonuniform manner, giving rise to a
    tree-like structure in space-time.}
\end{figure}
\begin{figure}[htbp]
  \subfigure[\label{f:shocks}]{\includegraphics[width=0.45\textwidth]
    {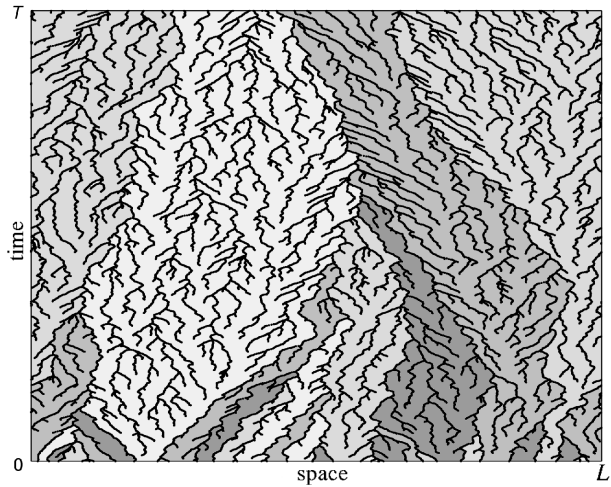}}\hfill
  \subfigure[\label{f:defTmini}]{\includegraphics[width=0.3\textwidth]
    {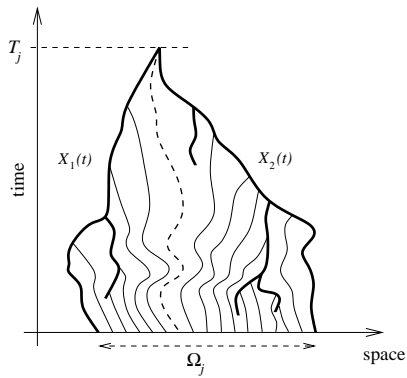}}
  \caption{(a) Shock trajectories for aspect ratio $L/L_f=32$ and with
    $T=10$; the different gray areas correspond to the space-time
    domains associated to the different smooth pieces $\Omega_j$'s of
    the velocity field at time $t=0$. (b) Sketch of the space-time
    evolution of a given smooth piece $\Omega_j$ located between two
    shock trajectories $X_1(t)$ and $X_2(t)$ which merge at time
    $T_j$. The trajectories of some fluid particles are represented;
    the shock trajectories are represented as bold lines; the dashed
    curve represents the trajectory of the unique fluid particle that
    is simultaneously absorbed by both shocks $X_1$ and $X_2$ at the
    time of their merger. This trajectory is a $T$-global minimizer
    if $T_j>T$.}
\end{figure}

In order to obtain a sketch of the behavior of the solution, the limit
of infinite aspect ratios $L/L_f$ is investigated numerically using a
kicked forcing and the fast Legendre transform method (see Sec.\
\ref{sec:periodic}). Moreover, we consider here the case of a forcing
whose spectrum is concentrated in the wavenumber intervals $|k|\in[\pi
L_f/L,\,3\pi L_f/L]$. As we will see in Sec.~\ref{sec:conclusion}, the
functional form of the forcing space correlation $B(\ell)$ at large
scale $\ell\gg L_f$ plays an important role. Numerical observations
suggest that, at any time in the statistical steady state (obtained
after sufficiently long integration), the shape of the velocity
profile is locally similar to the order-unity aspect ratio problem,
duplicated over independent intervals of size $L_f$ (see
Fig.~\ref{f:velocity}).  More particularly, when tracking the
trajectories of fluid particles backward in time, one can see that the
convergence of the minimizers to each other is far from uniform.
Figure~\ref{f:mini} shows that the minimizers are forming different
branches which are converging to each other backward in time, defining
in space time a tree-like structure.  Eventually, the number of
branches decreases and a unique trunk emerges around the global
minimizer. Note that the same type of behavior is also observed for
shocks (see Fig.~\ref{f:shocks}) by running forward in time.  The
intermediate time asymptotics correspond to the time scale for which
there is still a large number of such branches.

The velocity field at a given time, taken for convenience to be $t=0$,
consists of smooth pieces separated by shocks. Let us denote by
$\{\Omega_j\}$ the set of intervals in $[0,L)$, on which the solution
$u(\cdot,0)$ is smooth.  The boundaries of the $\Omega_j$'s are the
shock positions. Each of these shocks is associated to a root-like
structure formed by the trajectories of the various shocks which have
merged at times $t<0$ to form the shock under consideration (see
Fig.~\ref{f:shocks}). This root-like structure contains the whole
history of the shock and in particular its age.  Indeed, if the root
has a finite depth, the shock under consideration has only existed for
a finite time. A $T$-global shock is defined as a shock whose
associated root is deeper than $-T$. One can give the dual definition
for \emph{$T$-global minimizers}. All the smoothness intervals
$\Omega_j$ defined above, except that which contains the global
minimizer, will be entirely absorbed by shocks after a finite time
(see Fig.\ \ref{f:defTmini}).  For each of these pieces, one can
define a life-time $T_j$ as the time when the last fluid particle
contained in this piece at time $t=0$ enters a shock. It corresponds
to the first time for which the shock located on the left of this
smooth interval at time $t=0$ merges with the shock located on the
right. When the life-time of such an interval exceeds $T$, the
trajectory of this latter fluid particle is called a $T$-global
minimizer.

\begin{figure}[htbp]
  \centerline{\includegraphics[width=0.49\textwidth]
    {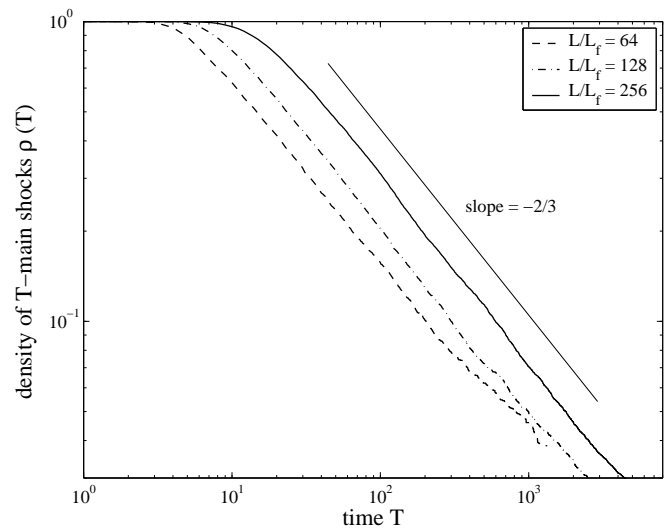}}
  \caption{Density of $T$-global shocks as a function of $T$ for three
    different system sizes $L/L_f=64$, $128$ and $256$; average over
    100 realizations.}
  \label{f:rho}
\end{figure}
Hence, for any age $T$, one can define a set of $T$-scale objects in
the spatial domain $[0,L)$ at a given time $t$. We define their
spatial density as the number of such objects divided by the size of
the domain $L$.  The density $\rho(T)$ of $T$-global shocks is
investigated numerically in the kicked case by using a two-step
method. Firstly, the simulation is run up to some large time $t$ for
which the statistically stationary regime is reached; secondly, each
shock present at time $t$ is tracked backward in time down to the
instant of its creation, giving an easy way to determine the density
$\rho(T)$.  We also average numerical results with respect to the
forcing realizations. It can be seen from Fig.~\ref{f:rho}, on which
the density for three different aspect ratios $L= /L_f$ is presented,
that the behavior of $\rho(T)$ is independent of $L$, and that it
displays a power-law $\rho(T)\propto T^{-2/3}$ in the intermediate
asymptotic range $L_f/u_{\rm rms} \ll T \ll L/u_{\rm rms}$.

We now present a simple phenomenological argument to explain the
scaling exponent $2/3$. Consider the solution at a given time ($t=0$,
for instance). Denote by $\ell(T)$ the typical spatial separation
scale for two nearest $T$-global shocks. Obviously, $\ell(T) \sim
1/{\rho(T)}$.  The mean velocity of a spatial segment of length
$\ell$ is given by
\begin{equation}
  b_\ell= \frac{1}{\ell}\int_{[y,\,y+\ell]}u(x,\,0)\,dx
  \label{minvelL}
\end{equation}
Since the expected value $\la u(x,\,0) \ra=0$ it is natural to assume
that for large $\ell$ one has the following asymptotics
\begin{equation}
  \int_{[y,\,y+\ell]}u(x,\,0)\,dx \sim \sqrt{\ell},
  \label{fluctmeanvelocity}
\end{equation}
which gives $b_\ell \sim \ell^{-1/2}$ for the mean velocity
fluctuations.  Consider now the rightmost minimizer corresponding to
the left $T$-global shock and the leftmost minimizer related to the
right one.  Since there are no $T$-global shocks in between, it
follows that the two minimizers we selected approach each other
backward in time for times of the order of $-T$. This means that the
backward-in-time displacement of a spatial segment of length $O(\ell)$
is itself $O(\ell)$ for time intervals of the order of $T$. The
corresponding displacement is given as the sum of two competing
effects: the first, a kind of drift induced by the local mean velocity
$b_\ell$, is connected with the mean velocity fluctuations and gives a
displacement $\propto b_\ell T$; the second effect is a standard
diffusive contribution $\propto T^{1/2}$ related to the diffusive
behavior of the minimizing trajectories. Taking both into account, we
get
\begin{equation}
\ell \sim C_1 T\ell^{-1/2} + C_2 T^{1/2},
\label{Tglob}
\end{equation}
where $C_1$ and $C_2$ are numerical constants. It is easy to see that
the dominant contribution comes from the first term. Indeed, if the
second term dominates, i.e.\ $\ell \propto T^{1/2}$, the first term
then gives a contribution of the order of $\ell^{3/2}$ which
contradicts to~(\ref{Tglob}). Hence, one should have $\ell \sim C_1
T\ell^{-1/2}$, leading to the scaling behavior
\begin{equation}
\ell(T) \propto T^{2/3}, \qquad \rho(T) \propto T^{-2/3}.
\label{scales}
\end{equation}
As we have already discussed above, $T$-global shocks are shocks older
than $T$. Denote by $p(A)$ the PDF for the age of shocks. More
precisely, $p(A)$ is a density in the stationary regime of a
probability distribution of the age $A(t)$ of a shock, say the nearest
to the origin. It follows from~(\ref {scales}) that the probability of
shocks whose age is larger than $A$ decays like $A^{-2/3}$, which
implies the following asymptotics for the PDF $p(A)$:
\begin{equation}
  p(A) \propto A^{-5/3}.
  \label{scales1}
\end{equation}

\section{Two related statistical models}
\label{sec:models}

\noindent We next discuss a very close analogy between the problem just
considered and the dynamical behavior of a simple first-order
stochastic ODE. Consider the following
one-dimensional stochastic Ito equation:
\begin{equation}
  dX=f(X)\,dW(t),
  \label{ito}
\end{equation}  
where $0<C_1\leq f(x)\leq C_2<\infty$, $x \in \rset$. We are
interested in the stochastic flow generated by~(\ref{ito}). In other
words, the aim is to understand the dynamical properties in the
large-time asymptotics of the trajectories corresponding to the
solutions of~(\ref{ito}) for different initial values. Let $X_1(t)$
and $X_2(t)$ with $t\geq 0$ be the two solutions associated to the
initial values $X_1(0)=x_1$ and $X_2(0)=x_2$. It is well-known
that the difference $[X_1(t) - X_2(t)]$ is a martingale which does not
change its sign; this implies that the limit
\begin{equation}
  c(x_1,x_2) \equiv \lim_{t \to \infty} \left [X_1(t) - X_2(t)\right]
  \label{lim}
\end{equation}  
exists and is finite.

\begin{figure}[htbp]
  \subfigure[\label{f:sketch-ODE}]{\includegraphics[width=0.49\textwidth]
    {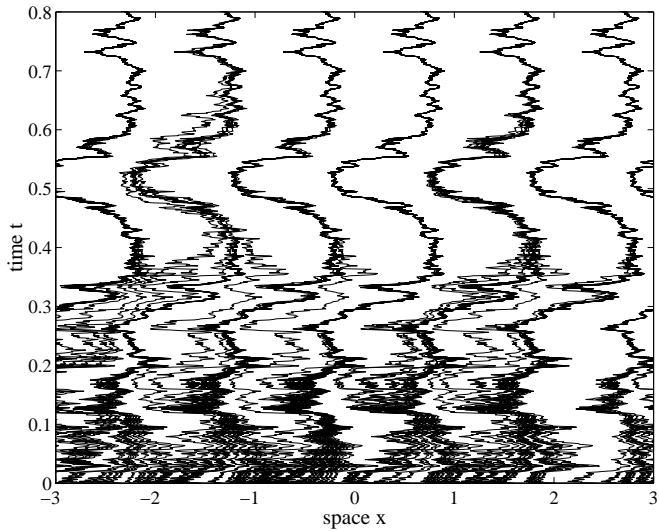}} \hfill
  \subfigure[\label{f:exponentiel-ODE}]{\includegraphics[width=0.49\textwidth]
    {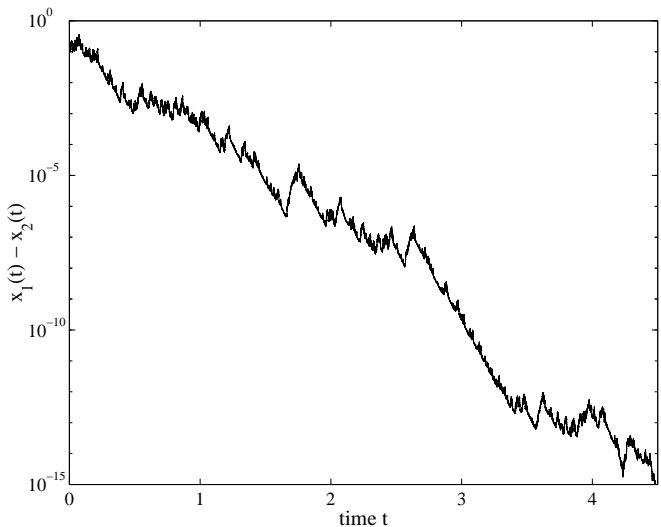}}
  \caption{Behavior of the solutions to the stochastic ODE (\ref{ito})
    for the periodic function $f(x)=2+\sin x$. (a) Unwrapped
    trajectories of the solutions converging to one of the periodic
    image. (b) Time behavior, for a given realization, of the
    separation between two trajectories $x_1(t)$ and $x_2(t)$
    associated to two different initial conditions. The two
    trajectories approach each other exponentially fast with a
    non-random rate $\lambda$.}
\end{figure}
Consider first the case when $f(x)$ is periodic. It is easy to see
that $c(x_1,x_2)$ has to be equal to a period of the function $f(x)$
(see, e.g., Fig.\ \ref{f:sketch-ODE}). Let, for simplicity, the
smallest period of $f(x)$ be equal to 1. Then, $c(x_1,x_2)$ can take
only two values: either 0 or 1.  In fact, it is possible to show (see
\cite{Bax1,Bax2,Bax3,Cal}) that the following picture holds. There
exists a random periodic sequence of points $x(k)=x_*(\omega) + k$, $k
\in \zset$, $x_* \in [0,1)$ such that $c(x_1,x_2)=0$ for all $x_1,x_2
\in (x_*(\omega) +k,\, x_*(\omega) + k +1)$ and $c(x_1,x_2)=1$ for
$x_1\in (x_*(\omega) + k -1,\, x_*(\omega) + k)$ and $x_2 \in
(x_*(\omega) + k,\, x_*(\omega) + k +1)$.  Moreover, in both cases,
the convergence is exponential:
\begin{equation}
  \left|c(x_1,x_2) - \left(X_1(t) - X_2(t)\right)\right| \sim
        C\exp(-\lambda t),
  \label{lyap}
\end{equation}
where $\lambda>0$ is a non-random Lyapunov exponent (see Fig.\
\ref{f:exponentiel-ODE}). The random variable $x_*(\omega)$ which
determines the positions of the separation points $x(k)$ depends on
the realization $\omega$ of the white noise $dW(t)$. If we consider
the stochastic flow on the unit circle, then all the trajectories are
stable except the one originating from $x_*(\omega)$. The stability is
governed by the non-random positive Lyapunov exponent $\lambda$. In
contrast, the trajectory which starts at $x_*(\omega)$ is unstable and
pushes nearby trajectories away. Hence, $x_*(\omega)$ plays a role
similar to that of the global minimizer for the Burgers equation with
a spatially periodic force. An exponentially small neighborhood of it
gets mapped into the whole unit circle apart from another
exponentially small interval which is the image of the rest of the
circle.  This instability island around $x_*(\omega)$ is similar to
the interval of fluid particles which are not absorbed by shocks until
time $t$.

All the results above in the case of periodic $f(x)$ have been
well-known for about 20 years and were presented here just to
emphasize an analogy with the main shocks and minimizers for Burgers
equation in a periodic situation. At the same time, the dynamical
behavior in the non-periodic case is more complicated and much less
studied. As we explain below, this behavior is also very similar to
the dynamical picture governed by the repelling $T$-global shocks in
the case of non-periodic Burgers equation.  Since $f(x)$ is
non-periodic $c(x_1,x_2)=0$ and all trajectories are asymptotic to
each other as $t \to \infty$.  However, as in the case of non-periodic
Burgers equation, this convergence is very non-uniform. Although there
are no globally unstable positions like $x(k)=x_*(\omega) + k$, for a
given time scale $T\gg1$, one still has exponentially small
$T$-instability islands which separate different intervals of
stability from each other. For a given time $T$, the image under the
stochastic flow of the $T$-instability islands covers almost all the
real line apart from exponentially small pieces which are images of
the stability intervals. Those exponentially small pieces, which we
call \emph{$T$-stability intervals} play a role similar to that of the
$T$-global shocks in Burgers turbulence.  For a larger time scale
$T_1\gg T$ a majority of $T$-instability island will become stable,
while some of them will still be unstable. Those islands are in fact
$T_1$-unstable.  One can ask the same question as studied above about
the asymptotics of the density of $T$-instability islands in the limit
as $T \to \infty$. Our preliminary numerical results indicate that the
density of $T$-instability islands scales as $T^{-1/2}$. Notice that
this scaling is different from the scaling $T^{-2/3}$ in the case of
Burgers equation. We believe that in the case of a stochastic flow the
$T$-instability islands and the density of them can be defined and
studied rigorously. The simplification is connected with the absence
of a slow drift which one has for Burgers equation due to fluctuations
of the average velocity. In the case of the stochastic flow the whole
asymptotic dynamical picture is formed by just two factors: diffusion
and hyperbolic contraction. As a result, the rigorous analysis of the
model seems to be a realistic task.

We finally suggest a simple statistical model which captures the
essential characteristics of the dynamics and the interactions of
$T$-global shocks and $T$-stability intervals. We believe that this
model is also quite interesting on its own.  Consider an infinite
system of particles on a one-dimensional integer lattice $\zset$.  Two
different particles cannot occupy the same site, and some sites are in
general not occupied. Each particle is associated to an integer age
$A\geq 0$. The model consists in describing a discrete time evolution
of particles. To get a configuration of particles at the next step, we
proceed as follows. First of all, with a probability $0<p<1$, we
generate independently new particles in every non-occupied site and we
set the age $A=0$ for all newly born particles.  Then, each particle
independently jumps, with a probability $1/2$, either to the right or
to the left by a distance $1/2$ and we add $1$ to the age of all
particles.  Notice that, after one time step, the particles are
located on the sites of the dual lattice. If there are two particles
which jumped into the same point of the dual lattice, we then assume
that the older one absorbs the younger, that is we ascribe to this
site a particle whose age is the maximum of the two ages of the
merging particles. It is easy to see that this dynamics reaches a
statistical steady state and one can define an integer-valued random
variable $A$ which is the age of the particle closest to the origin,
say from the right. The probability distribution of this random
variable, namely the asymptotics as $A \to \infty$ of the probability
$p(A)$, is closely connected with the statistics of $T$-instability
intervals for the non-periodic stochastic flow which we discussed
above. A numerical study of the model suggests that $p(A)$ scales like
$A^{-3/2}$ which corresponds to the scaling $T^{-1/2}$ for the density
of $T$-instability islands.  We postpone the detailed numerical and
theoretical analysis of the non-periodic stochastic flow and the
statistical model of interacting particles until a future publication.

\section{Concluding remarks}
\label{sec:conclusion}

\noindent We have studied the inviscid randomly forced Burgers
equation with non-periodic forcing on the whole real line started at
$t=-\infty$.  Our results indicate the existence of stationary regime
which corresponds to the velocities of one-sided minimizers and
suggest the following picture. At any given time $t$ (say $t=0$) and
any given $x \in \rset$, there exists at least one one-sided
minimizer.  However, due to fluctuations of the positions of one-sided
minimizers, there are no global minimizers. This means that any fluid
particle gets absorbed by a shock after a certain time.  Any two
one-sided minimizers are asymptotic to each other backwards-in-time,
that is the distance between them tends to zero as $t$ tends to
$-\infty$.  However, this convergence is very nonuniform. For fixed
$x,y \in \rset$ one-sided minimizers which originated at $t=0$ in $x$
and $y$ will approach each other exponentially fast beyond a certain
correlation time $T(x,y)$. This correlation time is of the order of
the maximum $T$ for which a $T$-global shock exists between $x$ and
$y$. One can say that $T$-global shocks form a hierarchical structure
separating different stability domains from each other. By stability
domains we understand intervals for which one-sided minimizers
converge exponentially fast with a correlation time of the order of
the turnover time $L_f/u_{\rm rms}$. The separation by $T$-global
shocks form separation ``walls'' between these different stability
intervals for times of the order of $-T$.  For larger backward times,
one-sided minimizers from the neighboring stability intervals are
exponentially asymptotic to each other.  Another interpretation of
$T$-global shocks is connected with their time of creation.  Every
shock can be traced backward only for a finite time interval.  For
$T$-global shocks, this time interval which determines the age of the
shock is larger than $T$.  Since all shocks have a finite age, it
follows that there are no true main shocks in the non-periodic
situation.  Our results suggest that the large-$T$ asymptotics of the
density of $T$-global shocks follows the power-law $\rho(T) \propto
T^{-2/3}$.  This gives a power-law behavior with exponent ${5}/{3}$
for the PDF of the age of shocks: $p(A)\propto A^{-5/3}$.  Another
related exponent is connected with the absorption times, defined in
the following way.  The absorption time $T(\ell)$ is the time after
which all fluid particles in the spatial interval $[-\ell/2,\,\ell/2]$
are absorbed by shocks.  Using the duality between the forward-in-time
behavior of fluid particles (minimizers) and the backward-in-time
behavior of shocks, we find that the absorption times $T(\ell) \propto
\ell^{3/2}$ at large $\ell$.

\begin{figure}[htbp]
  \centerline{\includegraphics[width=0.49\textwidth]{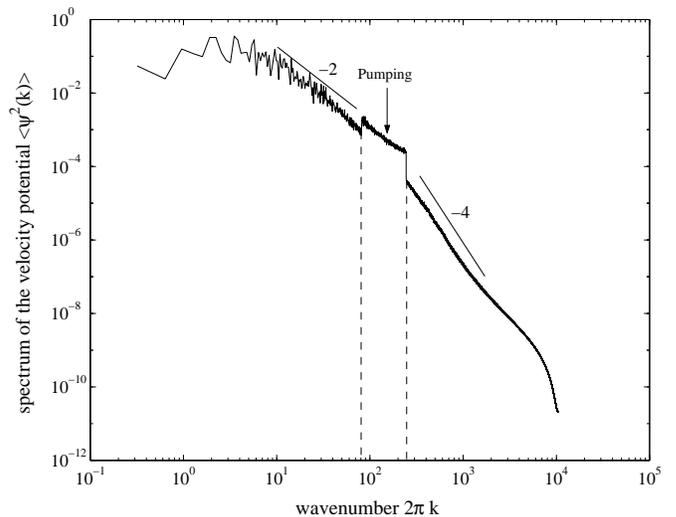}}
  \caption{Spectrum $\langle\hat\psi^2(k)\rangle$ of the velocity
  potential in the stationary regime for the aspect ratio
  $L/L_f=128$. This spectrum contains two power-law ranges: at
  wavenumbers $k\gg L/L_f$, the traditional $\propto k^{-4}$ inertial
  range connected to the presence of shocks in the solution and, for
  $k\ll L/L_f$, an ``inverse cascade'' $\propto k^{-2}$ associated to
  the large-scale fluctuations of $\psi$} \label{fig:spectrum}
\end{figure}
Notice that the power-law behavior of the density $\rho(T)$ of
$T$-global shocks can be interpreted in term of a kind of inverse
cascade in the spectrum of the solution (although there is no
conserved energy-like quantity). Indeed, the fluctuations
(\ref{fluctmeanvelocity}) of the mean velocity suggest that, for
large-enough separations $\ell$, the velocity potential increment
scales like
\begin{equation}
  \left | \psi(x+\ell,\,t)-\psi(x,\,t)\right| \propto \ell^{1/2}.
  \label{potincr}
\end{equation}
This behavior is responsible for the presence of an intermediate
power-law range with exponent $-2$ in the spectrum of the velocity
potential at wavenumbers smaller than the forcing scale (see Fig.\
\ref{fig:spectrum}). This power-law range corresponds for the velocity
spectrum to an equipartition of kinetic energy, meaning that the
large-domain asymptotics may be considered in the same universality
class as the KPZ problem for interface dynamics. It is also important
to notice that, in order to obtain this $-2$ range at small
wavenumbers, the spectrum of the forcing potential must go to zero
faster than $k^{-2}$ as $k\to0$.  Otherwise, the leading behavior is
non-universal and depends on the functional form of the forcing
correlation.

The randomly forced Burgers equation in an unbounded domain with a
different type of forcing was also considered in Ref.\ \cite{hk02},
where it was assumed that the forcing potential has at any time its
global maximum and its global minimum in a prescribed compact part of
the space. Such a forcing leads to a completely different behavior. In
particular, in this case there exists a global minimizer located in a
finite spatial interval for all times and all other minimizers are
asymptotic to it in the limit $t \to \infty$.

We finally stress that, in this paper, we have mostly discussed the
one-dimensional case.  It is of particular interest to analyze the
effects of non-compactness of the domain and the intermediate time
asymptotics in higher dimensions. The notion of main shock is then
replaced by that of topological shock \cite{ik01,bik02}, which are no
more isolated points, but spatially extended objects.  The natural
problem in this setting is to study geometrical and statistical
properties of the $T$-global shocks and to find the asymptotic
behavior of their age distribution.

\section*{Acknowledgments}
\noindent We are grateful to U.\ Frisch, H.\ Hoang, A.\ Kelbert, D.\
Khmelev, A.\ Kupiainen, R.\ Iturriaga and A.\ Sobolevski for
illuminating discussions and useful remarks. We are also grateful to
P.\ Baxendale who explained to us the behavior of the stochastic flows
generated by the first order ODE in the periodic case.  A big part of
this work was carried out in Vienna during the program on Developed
Turbulence in the Erwin Schroedinger Institute and we are very
grateful to the staff of the Institute for their warm
hospitality. This work was supported by the European Union under
contract HPRN-CT-2000-00162 and by the U.S.\ National Science
Foundation under agreement No.\ DMS-9729992. The numerical simulations
were performed in the framework of the SIVAM project of the
Observatoire de la C\^ote d'Azur, funded by CNRS and MENRT.

\end{document}